\documentclass[useAMS,usenatbib]{mn2e}
\usepackage{natbib}
\usepackage{url}
\usepackage{amssymb,amsmath}
\usepackage{txfonts,graphicx}

\newcommand\aj{{AJ}}%
\newcommand\apj{{ApJ}}%
\newcommand\apjs{{ApJS}}%
\newcommand\aap{{A\&A}}%
\newcommand\mnras{{MNRAS}}%
\newcommand\pasp{{PASP}}%
\newcommand\na{{New A}}%
\def\mapiii{MAPPINGS {\sc iii}}
\newcommand{\hii}{H\,{\sc ii}}
\newcommand{\ha}{\ensuremath{\mbox{H}\alpha}}
\newcommand{\hb}{\ensuremath{\mbox{H}\beta}}
\newcommand{\hg}{\ensuremath{\mbox{H}\gamma}}
\newcommand{\hd}{\ensuremath{\mbox{H}\delta}}
\newcommand{\oiii}{O\,{\sc iii}}
\newcommand{\nii}{N\,{\sc ii}}

\catcode`\@=11
\def\gapprox{\mathrel{\mathpalette\@versim>}}
\def\lapprox{\mathrel{\mathpalette\@versim<}}
\def\@versim#1#2{\lower2.45pt\vbox{\baselineskip0pt\lineskip0.9pt
      \ialign{$\m@th#1\hfil##\hfil$\crcr#2\crcr\sim\crcr}}}
\catcode`\@=12
\let\la=\lesssim            
\let\ga=\gtrsim 

\newcommand{\pccm}{\ensuremath{\,\mbox{cm}^{-3}}}

\title{The Balmer decrement of SDSS galaxies}
\author[B. Groves, J. Brinchmann, \& C.J. Walcher]{Brent Groves\thanks{brent@mpia.de}$^{1,2}$, Jarle Brinchmann$^{1}$, and Carl Jakob Walcher$^{3}$
\vspace*{6pt}\\
$^{1}$Leiden Observatory,  Leiden University, P.O. Box 9513, 2300 RA Leiden, The Netherlands\\
$^{2}$Max Planck Institute for Astronomy, K\"{o}nigstuhl 17, D-69117
Heidelberg, Germany\\
$^{3}$Astrophysikalisches Institut Potsdam, An der Sternwarte 16, D-14482 Potsdam, Germany}
\begin{document}
\maketitle
\begin{abstract}
High resolution spectra are necessary to distinguish and
correctly measure the Balmer emission lines due to the presence of strong metal
and Balmer absorption features in the stellar continuum. This accurate
measurement is necessary for use in emission line diagnostics, such as
the Balmer decrement (i.e.~\ha/\hb), used to determine the attenuation
of galaxies. Yet at high  
redshifts obtaining such spectra becomes costly. Balmer emission line
equivalent widths are much easier to measure, requiring
only low resolution spectra or even simple narrow band filters and
therefore shorter observation times. However a correction for the
stellar continuum is still needed for this equivalent width Balmer
decrement. We present here a statistical analysis of the Sloan Digital
Sky Survey Data Release 7 emission line galaxy sample, using the spectrally 
determined Balmer emission line fluxes and
equivalent widths. Using the large numbers of galaxies available in the SDSS 
catalogue, we determined an equivalent width
Balmer decrement including a statistically-based correction for the stellar continuum.
Based on this formula, the attenuation of galaxies can now be
obtained from low spectral resolution observations. 
In addition, this investigation also revealed an error in the \hb\ line fluxes,
within the SDSS DR7 MPA/JHU catalogue,  with the equivalent widths
underestimated by average $\sim$0.35\AA\ in the emission line galaxy
sample.  This error means that Balmer decrement determined
attenuations are overestimated by a systematic 0.1 magnitudes in $A_V$, 
and future analyses of this sample need to include this correction.

\end{abstract}


\section{Introduction}

The Balmer lines are the most well known and observed emission
lines in astronomy, being both strong lines in the optical and
ubiquitous as they arise from recombination to 
the $n=2$ level of the most common
element, hydrogen. As the atomic structure of hydrogen is so well
understood, the strength of the emission line fluxes can be well determined
if the radiation field ionizing the hydrogen gas is known, and,
importantly, the relative fluxes of the resulting hydrogen emission lines are
only weakly dependent on the local conditions. Given that the ratios are
well determined, the difference between the measured ratios of the Balmer lines and the
intrinsic values expected can be used to determine the reddening of
galaxies, or, more accurately, the ionized regions within them. In
association with a attenuation/reddening-law and selective-to-total attenuation, $R_{V}$, the
reddening can then give the total attenuation of a galaxy \citep[e.g.~
the oft used work of][]{Calzetti01}.  
This possibility of using the hydrogen emission lines, in particular
the ratio of the two strongest Balmer lines \ha\ and \hb, to measure
the reddening and attenuation has been known and utilized for many
years \citep[e.g.][ was one of the first mentions of the Balmer
decrement being affected by the path length through an absorbing
medium]{Berman36}.  

However, the accurate measurement of the Balmer decrement (i.e.~the ratio of
\ha/\hb) requires the measurement of both the relatively weak \hb\
line and the continuum underneath it to distinguish the
line. 
Separation of the underlying continuum from the Balmer emission lines is
vital as the  metal absorption lines and, especially,
Balmer absorption lines present in the spectrum of later-type stars
act to weaken or hide the relative flux of the emission lines in the
integrated spectra of galaxies. Such was shown by \citet{Liang04},
where the \hb\ line was not observed in a selection of galaxies in
low-resolution spectra ($R=150$) from the
the Canada-France-Redshift Survey, but moderate resolution spectra
($R>600$) revealed the weak \hb\ lines hidden by both dust and
absorption lines.
The exposure times needed to obtain sufficient spectral
resolution to distinguish the emission lines from the underlying stellar continuum
of galaxies means that studies of the Balmer decrement tend to be biased to
high emission-line equivalent width objects.  While this bias may not
be a serious issue, with dustier, more attenuated 
objects tending to have higher specific star formation rates and thus higher emission line
equivalent widths \citep[see e.g.][]{daCunha10}, such biases do tend to limit
samples for investigations of dust and star formation. This is
especially so at higher redshifts where the high spectral resolution
needed to resolve both the line and continuum limits surveys to the
brightest objects. 

Even with moderate resolution spectra, the Balmer
emission line fluxes are still sensitive to the way the stellar
absorption is accounted for, and this can be a
substantial source of error, particularly for weak emission line
sources.

It is these issues that motivates us to examine the possibility of
determining the decrement from the
\textit{equivalent widths} of the Balmer emission lines. 
Emission line equivalent widths do not require high resolution spectra to distinguish the line
from the continuum, allowing the use of low resolution spectra, such
as at R$\sim300$, enabling fainter objects or more objects to be observed  for the same
exposure time as high resolution spectra. Even at only
R$\sim100$, as will be available with the multi-object \emph{NIRSpec}
on JWST, the influence of \hb\ on the interpretation of galaxy spectra
can still be significant (Pacifici et al., in prep), and the line is
still detectable at a signal-to-noise $\sim5$ for 
some of the brighter emission-line galaxies.
The issue at these low resolutions (i.e.~$R<300$) becomes one of
distinguishing individual lines, such as [\nii]$\lambda6584$\AA\ from
\ha, not detecting the lines.

To examine how the Balmer line equivalent widths can be used to
determine the Balmer decrement we use the Sloan Digital Sky Survey \citep[SDSS][]{Abazaijian09},
which contains a large spectroscopic sample of emission line galaxies
covering a wide range of galaxy types and properties, including
attenuations. With such a wide range of galaxies, and reasonably high
resolution spectra ($R\sim1900$), the SDSS provides the perfect sample
for examining this issue.  With the large number of SDSS galaxies
it is possible to obtain a statistically representative estimate for the
correction factors needed (i.e. the
Balmer absorption lines) and thus determine the Balmer decrement from
equivalent widths alone. 

This method should be seen as complementary to the stellar spectral
synthesis continuum fitting methods widely used in the analysis of
galaxy spectra. Codes such as PLATEFIT \citep[][used for the
determination of the stellar properties in the SDSS MPA/JHU catalogue
used here]{Tremonti04} and \citep[][applied this to lower resolution,
higher redshift data in the VVDS sample]{Lamareille06}, pPXF
\citep{Cappellari04}, STARLIGHT
\citep{CidFernandes05}, STECKMAP \citep{Ockvirk06}, and VESPA
\citep{Tojeiro07}, all use linear combinations of synthetic stellar
population spectra \citep[such as from][]{Bruzual03} to fit the full observed
spectra of galaxies using various optimized maximum likelihood
approaches. These codes have been created to extract the maximum
possible information from galaxy spectra given degeneracies and noise
\citep[see e.g. the discussion in][]{Ockvirk06}, and thus are clearly
the best approach when strong continuum is detected. Yet the amount of
possible information to be extracted reduces with both decreasing signal-to-noise ratio
and spectral resolution. In addition, these methods are limited by the
available spectral libraries, which may not cover the full parameter
range needed to match the observed galaxies, and may have intrinsic
issues, as we demonstrate here in an issue we discovered in the course
of this paper. Thus an empirical method as we explore here is fully
complementary to the spectral synthesis methods used in most works.

In the following sections we introduce the Balmer
lines, both in absorption and emission (\S \ref{sec:balmer}), the SDSS emission-line galaxy
sample(\S \ref{sec:SDSSsample}, provide a possible way to determine the Balmer decrement from
the equivalent widths(\S \ref{sec:EWBalmer}), and finally also point out an interesting
problem with the fitting of the stellar continuum in the SDSS (\S \ref{sec:balmer}).

\section{The Balmer lines and Decrement}\label{sec:balmer}

\subsection{Balmer emission lines}

The Balmer emission lines in the interstellar medium arise
predominantly from the recombination and subsequent cascade of
electrons to the $n=2$ level of hydrogen. While collisional excitation
can also contribute to the Balmer line emission in hot media
\citep[see e.g.][]{Ferland09}, photoionization and recombination are
the predominant energetic processes in most galaxies.

As the atomic structure of hydrogen is so simple, it is possible to
determine the exact electronic transition rates, and therefore the ratios of resulting emission
lines from these transitions, as a function of physical conditions in
the interstellar  medium (see \citet[][]{Menzel37,Baker38} for the
original theory, with updated work by \citet{Seaton59} and \citet{Storey95}, and
treatments of this theory found in textbooks such as
\citet{Osterbrock06} or \citet{ADU}). In particular, two cases exist
for which the Balmer decrement has been determined over a range of
temperatures and densities: Case A and Case B. Case A assumes that an
ionized nebula is optically thin to all Lyman emission lines
(i.e.~lines emitted from transitions to the $n=1$ level of hydrogen), while
Case B assumes that a nebula is optically thick to all Lyman lines
greater than Ly$\alpha$ (i.e. transitions to $n=1$ from levels $n=3$
and above), meaning these photons are absorbed and re-emitted as a combination of
Ly$\alpha$ and higher order lines, such as the Balmer lines. These
two cases will lead to different intrinsic ratios for the Balmer
lines, with variations of the same order as temperature effects \citep[for
other possible ``Cases'' of emission which may occur, see e.g.][]{Ferland99,Luridiana99}. While
Case B is typically assumed for determining intrinsic ratios, in
reality the ratio in typical \hii\ regions lies between these two
cases, and must be determined using radiative transfer codes such as
\mapiii\ \citep[see eg][]{Groves04} or CLOUDY \citep{Ferland98}.  

\begin{table}
\begin{tabular}{lrrrr}
\hline
&$\lambda$ (\AA)&5000K& 10,000K & 20,000 K\\
\hline
\hline
\ha& 6562.80 & 3.04   & 2.86   & 2.75 \\
\hb& 4861.32 & 1.00  &  1.00   & 1.00 \\
\hg& 4340.46 & 0.458 & 0.468 & 0.475 \\
\hd& 4101.73 & 0.251 & 0.259 & 0.264 \\
\end{tabular} 
\caption{Balmer lines, including rest-frame wavelengths (Air), and their
  ratios relative to \hb\ for n$_e=10^2$\pccm\ and 3 different
  temperatures ([values from \citet{ADU}, based on data from \citet{Storey95}).}\label{tab:Balmer}
\end{table}

In Table \ref{tab:Balmer} we present the four strongest
Balmer emission lines and their ratios relative to \hb\ assuming Case B conditions. These
ratios are only weakly sensitive to density, with the \ha/\hb\ ratio at
$T=10^4$K equal to 2.86, 2.85, and 2.81 for the electron densities
n$_e=10^2$, $10^4$, and $10^6$ \pccm\ respectively, hence we only show
the larger variation due to temperature here. For a full ratio
description see Table B.7 in \citet{ADU}, or Table 4.4 in
\citet{Osterbrock06}. While these variations due to temperature and
density are significant, they are still small relative to the effects
of dust, as visible in the later sections, and hence strong
diagnostics for the amount of reddening experienced by an emission
line galaxy.

\subsection{Balmer absorption lines}

As discussed in the introduction, the fitting of the underlying
stellar continuum is a vital step in determining line fluxes,
especially for hydrogen (e.g.~Balmer) and helium recombination lines, which have underlying
corresponding absorption lines. 
The Balmer absorption lines arise from the absorption of light by
hydrogen in the excited $n=2$ level in the photospheres of stars. The
strength of the absorption is dependent on both the effective temperature
and gravity, as they require a significant  fraction of hydrogen to be
excited to the $n=2$ level \citep[see e.g., Table 4 in ][]{GD99a}. The
maximum equivalent widths occur around T$_{eff}\sim 9000$K,
corresponding to early A-type stars.

\begin{figure}
\includegraphics[width=\hsize]{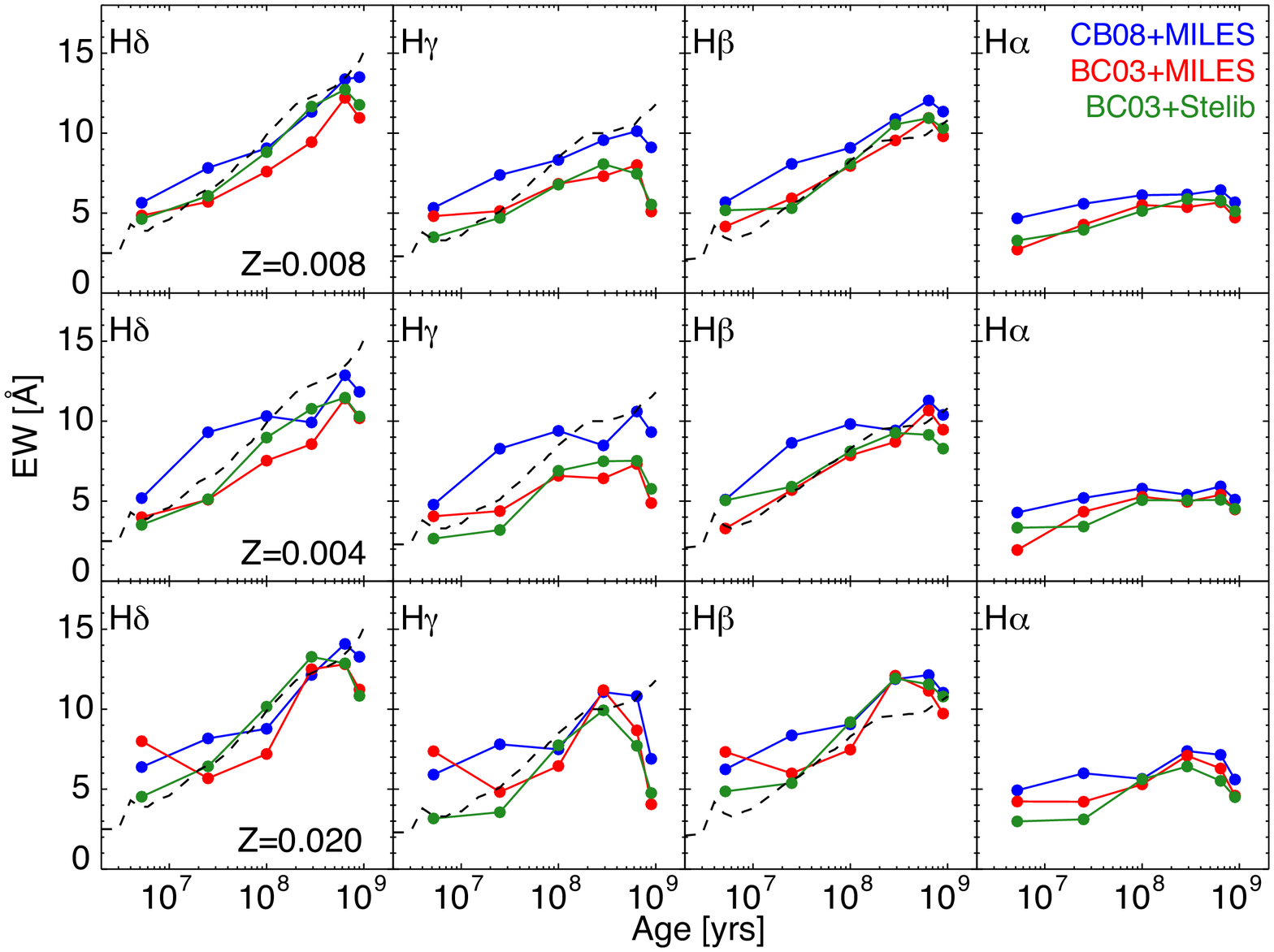}
\caption{Equivalent widths (EW) of the four dominant Balmer lines,
  \ha, \hb, \hg, and \hd\ (as labelled), for simple stellar populations of varying
  age and metallicity (increasing from top to bottom row, as marked on
  lower left in \hd\ figure). Three
 stellar synthesis models are considered, as indicated by the colours
 in the key in the upper right; the \citet[][BC03]{ Bruzual03} stellar
 population synthesis code using the MILES 
\citep{SanchezBlazquez06,Cenarro07} and Stelib \citep{LeBorgne03}
stellar spectra libraries, and the 2008 
version of the Charlot \& Bruzual (in prep, CB08) code with the MILES
library. The dashed line shows the
results from \citet{GD99a}. }\label{fig:EW_Balmer}
\end{figure}

For a simple, single-aged stellar population, the equivalent widths
(EW) of the Balmer stellar absorption features depend on the age and,
more weakly, on the metallicity, and vary from $\sim2$\AA\ to 
$\sim15$\AA\, with the maximum value occurring for stars aged around 500 Myr
(i.e.~dominated by the light from A \& F stars). 
In Figure \ref{fig:EW_Balmer} we show the
variation of the equivalent widths (EW) respectively for simple stellar populations as a function of
age and metallicity (as labelled in the left hand of the figure in the \hd\ diagrams). We compare the
four strongest lines; \ha, \hb, \hg, and \hd, using four models to
determine the equivalent widths as labelled in the upper right; the
\citet[][BC03]{ Bruzual03} stellar population synthesis code using the MILES
\citep{SanchezBlazquez06,Cenarro07} and Stelib \citep{LeBorgne03}
stellar spectral libraries, and the 2008 
version of the Charlot \& Bruzual (in prep, CB08) code with the MILES
library.  Also shown by the dashed
lines are the results from \citet{GD99a}. The weak dependance on
metallicity and the strong dependance on age, with the peak in EW at
$0.5-1$Gyr for all lines, are clearly seen for all models. The equivalent widths are 
reasonably similar between the dominant Balmer lines (i.e.~\ha, \hb,
\hg, and \hd)  varying at most a factor of $\sim2$
\citep{Kauffmann03a,GD99a,GD99b}. 

Thus, as the equivalent widths of the
stellar absorption features are approximately constant with wavelength
while the relative strength of the Balmer emission lines decrease
rapidly with decreasing wavelength (i.e.\ for the higher order lines),
stellar absorption affects strongly the measurement of the Balmer
decrement i.e.~\ha/\hb, \hg/\hd. This is especially so for weak
emission line galaxies where the stellar absorption features are
relatively stronger and only \ha\ is seen in emission.
This relative importance of the effect of the stellar Balmer absorption on the
emission lines is important when considering the Balmer ratios, as
discussed in later sections.

\section{The SDSS Sample}\label{sec:SDSSsample}

Within this work we base our findings on the spectroscopic data from the  seventh Data Release of the SDSS
\citep[DR7][]{Abazaijian09},  though we also refer to the fourth Data
Release \citep[DR4][]{DR4} as well when necessary. The SDSS used a
pair of multi-fibre spectrographs with fibres of 3" diameter. In
most galaxies the fibres were placed as close as possible to the
centres of the target galaxies.  The flux- and wavelength-calibrated 
spectra cover the range from 3800 to 9200\AA, with a resolution of
$R \sim$1900. 

We obtain our emission line fluxes from the MPA/JHU analysis of the
SDSS spectroscopic sample\footnote{The data catalogues are available
  from \url{http://www.mpa-garching.mpg.de/SDSS/}}. This database
contains, in addition to the emission line fluxes, derived physical
properties for all spectroscopically observed galaxies in the SDSS
DR7. The procedure for emission line measurement, detailed in
\citet{Tremonti04}, was to correct the line fluxes for stellar
absorption, fitting a non-negative combination of stellar population
synthesis models from Charlot \& Bruzual (in prep., CB08)\footnote{The
  model spectra used were from an early version of the models and
  differ from what will be eventually published. The differences from
  BC03 are primarily due to different treatment of TP-AGB stars and
  that the empirical stellar library used was the MILES library rather
  than STELIB.} for the 
SDSS DR7 release and \citet[][BC03]{Bruzual03}  for the SDSS DR4 release
\footnote{The spectra are available as part of the GALAXEV package,
  which is can be obtained from
  \url{http://www2.iap.fr/users/charlot/bc2003/index.html}.}.
The best-fitting stellar population model
also places constraints on the 
star formation history and metallicity of the galaxy \citep[see
e.g.][]{Gallazzi05}, and has been used to estimate stellar masses and
star-formation histories \citep{Kauffmann03a}. 

The equivalent widths of the Balmer lines we use here, also available
as part of the MPA/JHU database (listed as \texttt{(line name)\_reqw}
in the \texttt{gal\_line} data file), are computed
from straight integration over the continuum-subtracted bandpasses
listed in table \ref{tab:EWbands}.  Note that, by definition, emission
lines have negative values of equivalent width but for clarity in the
rest of the paper we
assign all emission lines a positive value. 
The continuum in this case is estimated using a running median with a
200 pixel window and does not properly account for stellar 
absorption. This measurement is representative of the cases where a
more accurate determination of the stellar continuum, as done with the
MPA/JHU database, is not possible, such as with low resolution or low
S/N data, and helps characterize the effects of stellar absorption on the lines.

\begin{table}
\caption{Equivalent width bandpass}\label{tab:EWbands}
\begin{tabular}{|l|c|c|c|}
\hline
Line & Centre (\AA) & Lower bound (\AA) & Upper bound (\AA) \\
\hline
\hd & 4101.73 & 4092.0 & 4111.0 \\
\hg & 4340.46 & 4330.0 & 4350.0 \\
\hb & 4861.32 & 4851.0 & 4871.0 \\
\ha & 6562.80 & 6553.0 & 6573.0 \\
\hline
\end{tabular}
\end{table}

\begin{figure}
\includegraphics[width=\hsize]{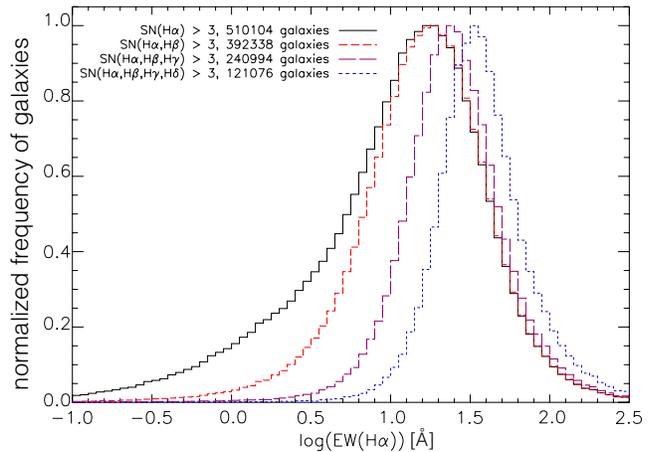}
\caption{Distribution of the \ha\ emission line equivalent widths for
  the samples considered in this work. All four histograms are
  normalized to their peak value, with the total number of galaxies in
  each sample listed in the key in the upper left. The solid histogram
  shows all emission line
  galaxies in SDSS DR7 (defined by the presence of \ha\ in emission),
  and the three dashed curves show the distributions of samples
  defined by S/N cuts in the Balmer lines (as labelled in the upper left)
  considered in this work. }\label{fig:EWHa_wcuts}
\end{figure}

As we concentrate on emission line galaxies in this work, in
particular galaxies with measurable Balmer
emission lines, we have placed cuts on the signal-to-noise
(S/N) of the Balmer emission lines using the uncertainties given by the MPA/JHU
catalogue. As discussed on the website, the listed uncertainties are
formal, and likely underestimates, thus we increase the uncertainty estimates on the
emission lines to take into account continuum subtraction
uncertainties by the factors listed on the web site determined by
comparisons of duplicate observations within the SDSS sample.  Specifically for the Balmer
lines, we multiply the line flux
uncertainty estimates by a factor of 1.882.  
From the full SDSS emission line galaxy sample we define three galaxy
samples depending on the 
lines and ratios being examined; SN(\ha,\hb),
SN(\ha,\hb,\hg), and SN(\ha,\hb,\hg,\hd), where we require a S/N$>3$
in two or more of the four strongest Balmer lines (\ha, \hb, \hg, \& \hd).
The S/N cuts are dominated by the weakest line in each sample due to the strong decrease
in relative flux for the higher order lines, thus the inclusion of each
higher order line biases the samples to higher equivalent widths of the \ha\ emission
line. 
As shown in figure \ref{fig:EWHa_wcuts}, beginning from the full
galaxy sample of SDSS DR7 ($\sim$ 928,000 galaxies), approximately
half are emission line galaxies ($\sim$ 510,000, as measured by the
presence of \ha\ in emission), with a broad spread of equivalent widths peaking at
$\sim20$\AA\ (as measured from the local continuum, not corrected for
stellar absorption). As each higher order Balmer line is included, the
sample rapidly decreases and is biased to higher equivalent widths,
with the SN(\ha,\hb,\hg,\hd) sample limited to $\sim$120,000 galaxies
with EW(\ha)$>10$\AA, with a distribution peaking at 32\AA\
(SN(\ha,\hb) has $\sim$392,000 
galaxies, while SN(\ha,\hb,\hg) has $\sim$241,000 galaxies). As the
emission line EW(\ha) in a galaxy
spectrum can be considered a proxy for the specific star formation rate
(the current star formation rate relative to the total stellar mass,
sSFR=SFR/M$_*$) of a galaxy, the bias in EW(\ha) means a bias to more 
``starforming'' galaxies, which means a bias to lower-mass,
lower-metallicity, bluer galaxies as shown in previous works
\citep{Brinchmann04,Tremonti04}. This bias needs to be kept in mind when
considering the diagrams and analysis in this work.  

Figure \ref{fig:EWHa_wcuts} also reveals the limitation of
low-resolution spectroscopy in finding all emission line 
sources, and thus the limitation of applicability of the method we
explore here. Approximately 5\% of the SN(\ha) sample, and even 0.7\% of
the SN(\ha,\hb) sample, actually have EW(\ha) less than zero (i.e.~the
emission line is lost in the stellar absorption feature). These
sources would never be picked up as emission line galaxies in
low-resolution spectra, and the use of an emission-line equivalent
width Balmer decrement to determine the attenuation would return
spurious results.

Note that we have not included any cuts on redshift or the type of
emission-line galaxy as in previous works on SDSS emission line galaxies
\citep[e.g.][]{Kewley06}. Such redshift cuts are necessary to make certain that
aperture effects do not play a part in the derived galaxy properties
\citep[by sampling only a small, biased part of the galaxy, as
discussed in][]{Kewley05} and to prevent
luminosity biases at the higher redshifts. Yet as we care only for the
derived Balmer emission line fluxes and stellar equivalent widths
these issues do not strongly affect our findings.
 
Separating emission-line galaxies by class (i.e.~star-forming or
Active Galactic Nucleus dominated)
is necessary when examining the structure of the forbidden emission
lines (e.g. [\oiii]$\lambda 5007$\AA) which depend strongly on the
dominant ionization mechanism in the gas \citep[see e.g.][]{Kewley06},
and when comparing physical galaxy properties with emission line
properties \citep[see e.g.][]{Kauffmann03a,Kauffmann03b}. However, as
discussed in the previous section, the relative strength of the
Balmer lines depend only weakly on local conditions, and will vary little in their intrinsic ratios
between being photoionized by AGN or by OB stars (i.e.~\ha/\hb\ should be $\sim
2.86$ in star-forming galaxies and $\sim 3.1$ in galaxies ionized
purely by an AGN). Thus, while this difference is significant and will
have some bearing on the work in this paper, it is secondary to the
effects of dust attenuation.  
Only when collisional heating dominates
the atomic gas and the excitation of hydrogen, such as in shocks or
clouds in hot gas, can the intrinsic ratios be significantly different
\citep[see e.g.][]{Ferland09}, but these processes are not expected to
dominate most galaxies within our sample. An additional reason to
include AGN is that, in accordance
with our main aim, separating out
the contribution of AGN from lower S/N samples may prove problematic
as the weak diagnostic lines become lost within the noise.

Approximately 2\% of the full SDSS sample considered here are
duplicate observations of galaxies (i.e.~$\sim4$\% of the sample are pairs).
We have not removed these from the sample so as to include the
intrinsic scatter due to observational uncertainties in the
subsequent analysis.

\section{The Equivalent Width Balmer Decrement}\label{sec:EWBalmer}

\begin{figure}
\includegraphics[width=0.48\hsize]{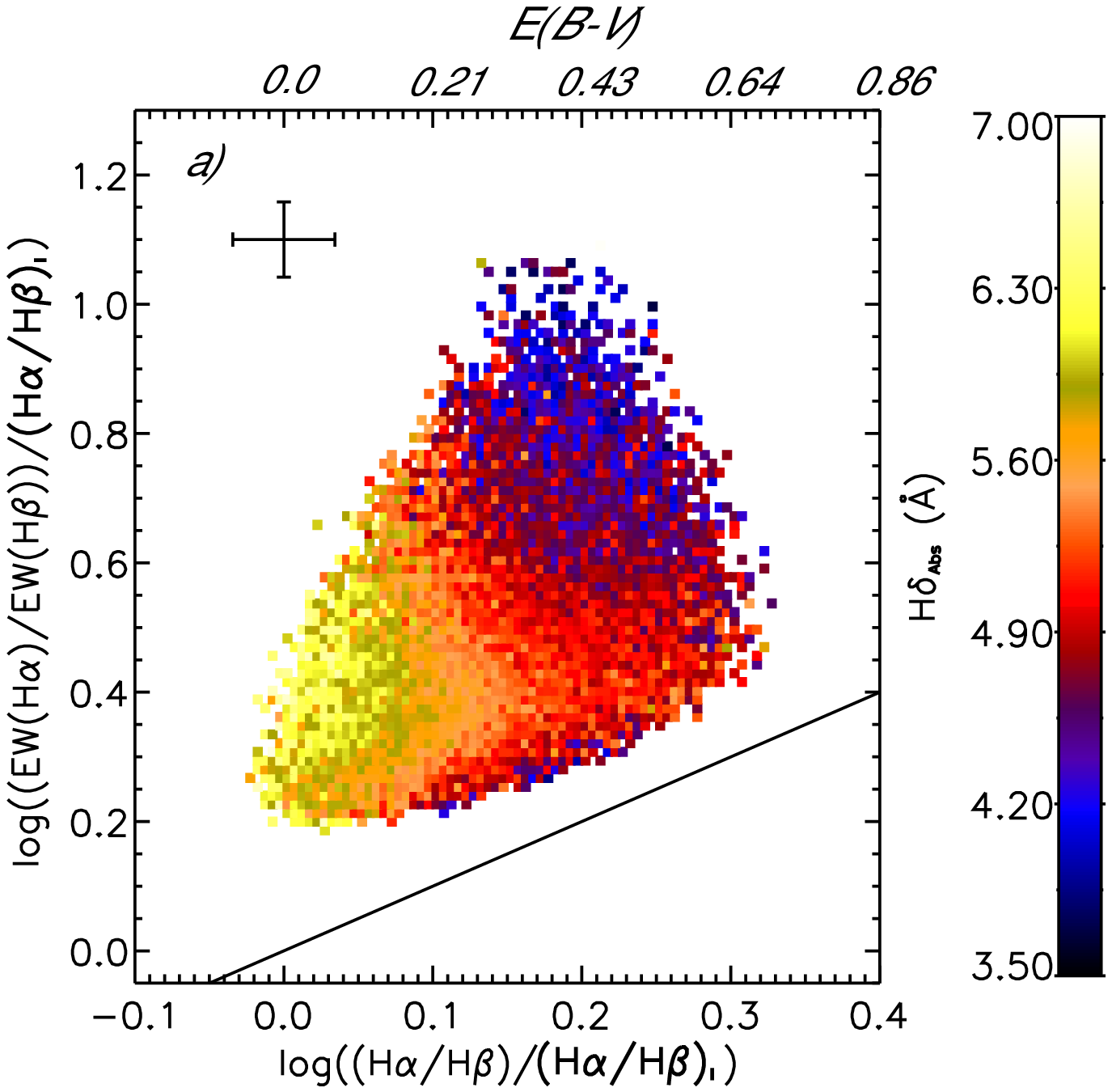}
\includegraphics[width=0.48\hsize]{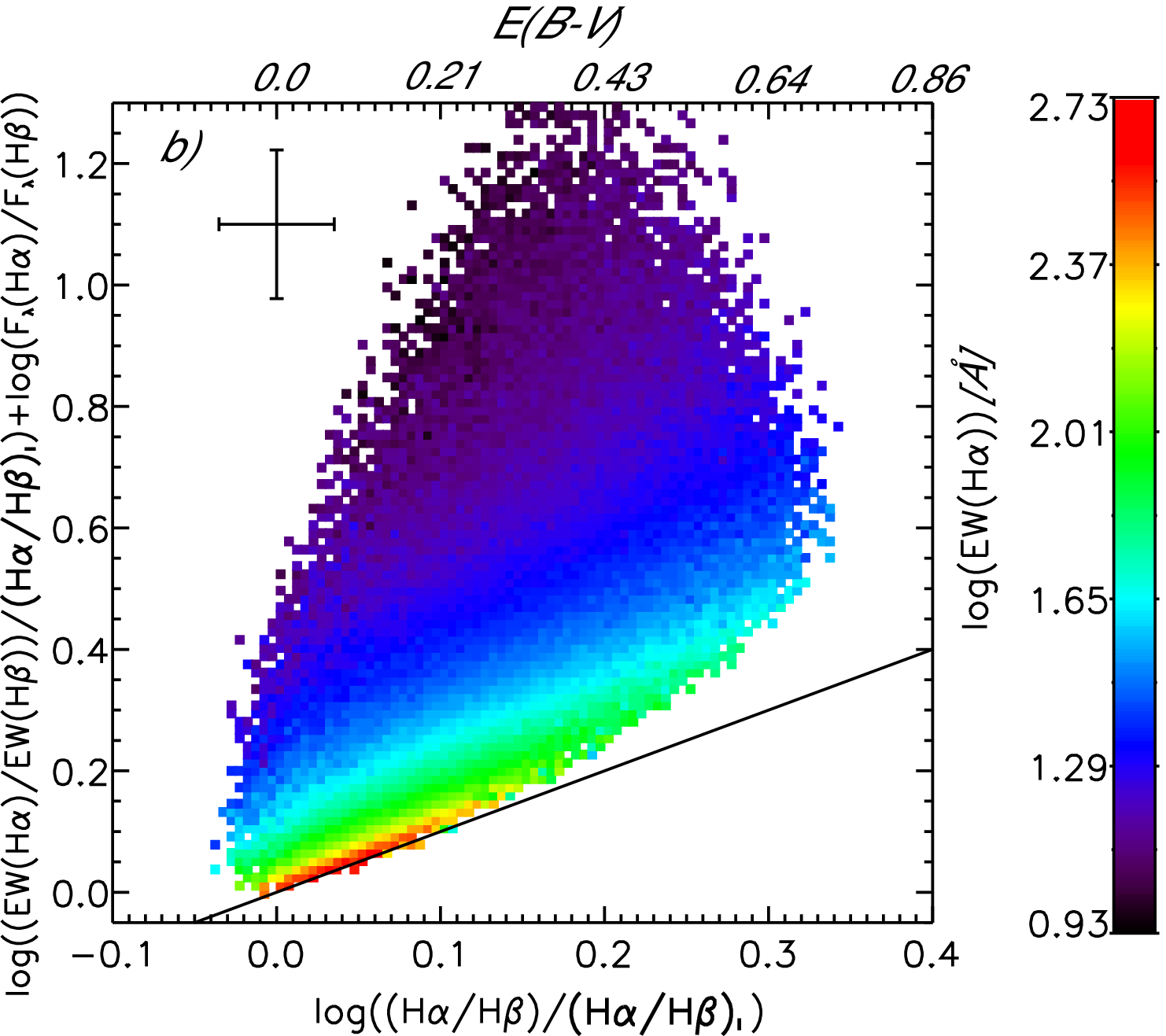}
\caption{The variation of the equivalent width based Balmer decrement
($\log[$EW(\ha)/EW(\hb)$]$) versus the stellar
continuum-subtracted flux based Balmer decrement for the SDSS
SN(\ha,\hb) sample. All axes are normalized by the intrinsic Balmer
ratio, $(\ha/\hb)_{\rm I}=2.86$. In the left diagram, the EW(\ha/\hb) has not been
corrected for the difference in continuum flux at the \ha\ and \hb\
wavelengths (${\rm F}_{\lambda}(\ha)/{\rm F}_{\lambda}(\hb)$), while
in the right this is included. The upper left hand corner shows the
median uncertainties for the sample, and the straight line indicates a
1:1 relation. The colours and associated colourbars
indicate the median $\hd_{\rm Abs}$ and log[EW(\ha)]
respectively in each pixel. The top axes give the resulting E$(B-V)$
from the balmer decrement assuming the \citet{ODonnell94} Galactic extinction
curve.}\label{fig:EW_HaHb} 
\end{figure}

The issue of simply using directly the equivalent widths of the Balmer lines as proxies
for the line fluxes when calculating the Balmer decrement can be seen
in figure \ref{fig:EW_HaHb} which shows the spread of the equivalent
width based Balmer decrement ($\log[$EW(\ha)/EW(\hb)$]$) against the
``true'' Balmer decrement determined  from the stellar
continuum-subtracted line fluxes (log(\ha/\hb)) for the SN(\ha,\hb)
sample. Both the EW Balmer decrement and the flux Balmer decrement
have been normalised to the intrinsic ratio of 2.86, appropriate for a
low density gas of $T=10^4$\,K. 

One of the first obvious issues to be corrected for is the variation
of the underlying stellar continuum between the \ha\ and \hb\
wavelengths. In the left diagram of figure \ref{fig:EW_HaHb}, we show
the distribution of $\log[$EW(\ha)/EW(\hb)$]$, uncorrected for the
continuum flux variation, while on the right the more accurate form of
the EW Balmer decrement is used:  $\log[$EW(\ha)/EW(\hb)$] +
\log[{\rm F}_{\lambda}(\ha)/{\rm F}_{\lambda}(\hb)]$, where
${\rm F}_{\lambda}(\ha)$ is the continuum flux at \ha, determined from
a 200 pixel median smoothing of the emission-line subtracted continuum.

When uncorrected for the underlying continuum variation there is a
clear systematic offset of the EW Balmer decrement from the 1:1 relation of $\sim
0.1$ dex. Correcting for the continuum variation removes this offset,
yet a significant spread remains. This spread is due to the effect of the
stellar Balmer absorption features. Without these,
EW(\ha)$\times{\rm F}_{\lambda}(\ha)$ should be, by definition, the flux of the
line. The colors in figure \ref{fig:EW_HaHb}a
indicate the median \hd\ absorption index \citep[\hd$_{\rm
  Abs}$,][]{Worthey97,Kauffmann03a} of the sample in each pixel. As
the figure shows, while the absolute strength of the stellar Balmer absorption
features does play a part in
the observed offset of the SDSS galaxies' EW Balmer decrements,
the dominant mechanism for the offset and spread is the
\emph{relative} strength 
of the stellar absorption features to the emission lines. This can be
seen by the distribution of 
the equivalent width of the \ha\ emission line indicated by the
colours in figure \ref{fig:EW_HaHb}b, where there is a clear gradient
of decreasing EW(\ha) with increasing offset from the line. As the
emission lines become weaker overall, the stellar absorption features, which
are $<$10\AA\ as discussed in section \ref{sec:balmer}, obscure
a greater fraction of \hb\ relative to \ha\ and therefore lead to a larger offset.

As discussed in the introduction, the best way to compensate for the
effect of the stellar absorption features on the emission lines is to
fit the stellar continuum as done within the MPA/JHU SDSS
database. However, when only poor quality spectra are available such
as for high redshift galaxies, the determination of the Balmer
absorption features may be unreliable.  One possible approach when
faced with low resolution spectra is to assume that the absorption
equivalent width is constant for both Balmer lines across the whole
sample. While figure \ref{fig:EW_Balmer} clearly shows that the
absorption EW is \emph{not} the same for all the Balmer lines, it
provides a first step when information is sparse and uncertainties
large. When a constant Balmer absorption correction $R$ is assumed for
both EW(\ha) and EW(\hb), a correction factor of $R=4$\AA\ is
determined when the offset of the SDSS galaxies' EW Balmer decrements
to the measured \ha/\hb\ ratios is minimized using an error-based
weighting. The inclusion of this simple correction factor improves the
situation when compared to that shown in figure \ref{fig:EW_HaHb}, but
with a still
significant scatter of $\sigma\sim0.1$ dex around the expected value and
an extended tail of objects
towards lower values. Both the scatter and the tail arise due to the
assumption of a constant offset (i.e.~Balmer absorption) for the whole
sample. The value determined is biased towards high EW(\ha) galaxies,
as these galaxies both dominate the sample and have lower
uncertainties, as discussed in section \ref{sec:SDSSsample}.

%
%

\begin{figure}
\includegraphics[width=\hsize]{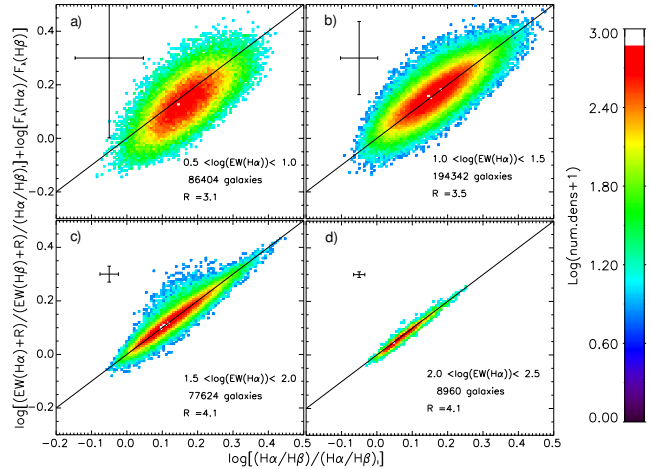}
\caption{2D histograms of the distribution of the SDSS
SN(\ha,\hb) sample galaxies' equivalent-width based Balmer decrements
($\log[$EW(\ha)/EW(\hb)$]$), including a constant correction for stellar
Balmer absorption, versus the stellar continuum-subtracted flux based Balmer
decrements. Each figure shows a different bin  of \ha\ equivalent
width, as labelled in lower right of each 
plot (in \AA). As in figure \ref{fig:EW_HaHb}, both axes are normalized by the intrinsic Balmer
ratio, $(\ha/\hb)_{\rm I}=2.86$. The colours indicate the log of the number density in each
pixel, as labelled by the colour bar on the right, with pixels with
less than 5 galaxies excluded. The error bars in the upper left indicate the median
uncertainty for each sample, with the total number of galaxies listed
in the lower right of each plot. 
Note that the $y$-axis has a different correction for stellar Balmer
absorption, $R$, for each plot as indicated in the lower left of each
plot, and that galaxies with \hb\ still in absorption
after correction (i.e.~${\rm EW}(\hb)+R<0$) have been excluded from the
sample.}\label{fig:binEWfix}
\end{figure}

When split into bins of different EW(\ha), more accurate
fits with differing correction factors are obtained. In figure
\ref{fig:binEWfix}, we show the fits for the galaxies split into four
bins; $0.5<\log({\rm EW}(\ha))<1.0$, $1.0<\log({\rm EW}(\ha))<1.5$,
$1.5<\log({\rm EW}(\ha))<2.0$, and $2.0<\log({\rm EW}(\ha))<2.5$.
The number of galaxies in each bin is listed in the lower right of
each figure, and the median uncertainties are shown by the error bars
in the upper left. Note that galaxies with \hb\ still in absorption
after  the correction factor $R$ is applied
have been excluded, but that this is less than $0.2$\%\ of the sample
in each bin. 
As can be seen in the figure, the median uncertainties increase
quickly with
decreasing emission line equivalent width. It is for this reason that
galaxies with $\log({\rm EW}(\ha))<0.5$ have not been included here. 

The $y$-axis for all four plots includes a correction for stellar
absorption; $\log([$EW(\ha$)+R]/[$EW(\hb$)+R]) +
\log[{\rm F}_{\lambda}(\ha)/{\rm F}_{\lambda}(\hb)]$. 
As for the full sample, we determine the correction factor, $R$, for each binned sample
 by finding the value that leads to the minimum offset from the 1:1
 relation with the uncertainties giving
 $1\sigma$ offsets from this relation. The values determined are $R=
 3.1$, $3.5$, $4.1$, and $4.1$\AA\ for each increasing bin of EW
respectively, as indicated in the lower-left of each plot in figure
\ref{fig:binEWfix}. The $1\sigma$ uncertainty around $R$ for each bin
is approximately 1.0 (slightly less for the EW(\ha)$>100$\AA\
bin). The lowest bin has an uncertainty of ~1.5, though the
probability distribution for $R$ is slightly skewed to higher values,
arising from the offset visible in \ref{fig:binEWfix}a, discussed
below.
The Balmer decrement determined from the EWs is significantly better
than for the full sample, and especially so when no correction is
included (figure
\ref{fig:EW_HaHb}), with dispersions of
$1\sigma\sim$0.11, 0.06, 0.05, and 0.04 dex around the 1:1 line
respectively for the 4 binned samples. 

The lowest EW(\ha) (figure
\ref{fig:binEWfix}a) sample appears by eye to be slightly offset form
the line. A reasonable hypothesis is that this offset arises due to
our simple assumption of a constant correction $R$ to both EW(\ha) and
EW(\hb), whereas it is clear from \ref{fig:EW_Balmer} that the
absorption EW(\ha) is typically less than the absorption EW of \hb\ by
approximately a factor of 0.6 on average. However, changing the
correction factor of EW(\ha) to $0.6R$ and redoing the fit does not
remove this offset. On closer examination it is clear that this offset
is due to a biasing of the fit to the high 
signal-to-noise data, which predominantly occur at high values of the
\ha/\hb\ ratio due to measurement biases (i.e.~ there is a clear
gradient in EW(\ha) from top to bottom in figure
\ref{fig:binEWfix}a). Assuming uniform weighting for the fit
(i.e.~ignoring the errors in EWs) 
gives a value of $R=3.0$, well within the large uncertainties for
$R$. For the other figures, assuming an offset correction factor for
EW(\ha) of $0.6R$, results in $R=3.5$, $3.7$, and $3.7$\AA\ in terms
of increasing EWs, with similar dispersion around the relations. The
results are within the uncertainties for $R$ when assuming a constant
correction, but in all cases indicate the necessity for the correction
of stellar absorption to the emission lines of a factor of $\sim4$\AA.

For figures \ref{fig:binEWfix}a, b, and d ($0.5<\log({\rm
  EW}(\ha))<1.0$, $1.0<\log({\rm EW}(\ha))<1.5$, and $2.0<\log({\rm
  EW}(\ha))<2.5$ respectively) the observed scatter is less than the
median uncertainty. Only for figure  \ref{fig:binEWfix}c does there
appear to be a significant, low number scatter above the line (note
the log scale density in figure  \ref{fig:binEWfix}). However, when
examined, the scatter in this diagram, and also in figures
\ref{fig:binEWfix}a, b, and d, is correlated with the uncertainty in
the determined \ha\ and \hb\ lines, with the median scatter in each
pixel increasing
significantly the further from the line. Thus the median uncertainty
for the outliers is greater than the median uncertainty of the sample
as a whole in figure \ref{fig:binEWfix}c.

\begin{figure}
\includegraphics[width=\hsize]{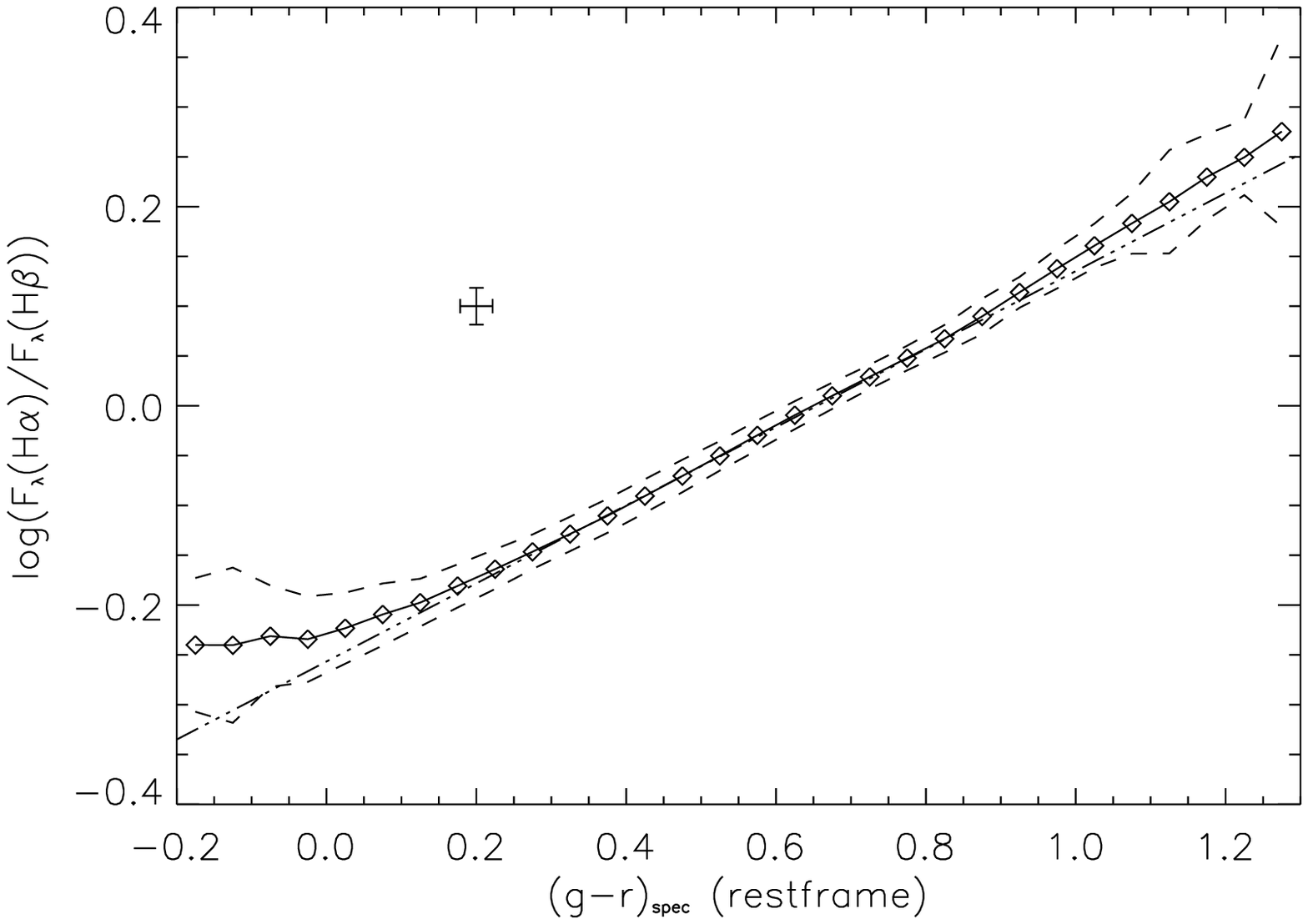}
\caption{Distribution of the full SN(\ha,\hb) SDSS sample ratio of
 continuum fluxes measured at the \ha\ and \hb\ wavelengths
 (${\rm F}_{\lambda}(\ha)/{\rm F}_{\lambda}(\hb)$) against the restframe
 $g-r$ colour as measured from the SDSS fibre spectrum. The solid line
 with diamonds shows the median ${\rm F}_{\lambda}(\ha)/{\rm
   F}_{\lambda}(\hb)$, while the dashed lines show the 1$\sigma$
 dispersion. Median uncertainties are indicated by the error bars in the upper left. The
 thick dot-dot-dashed line indicates the best fit linear relation, with
 $y=-0.26+0.39(g-r)$ .}\label{fig:Bcont_colour} 
\end{figure}

While figure \ref{fig:binEWfix} demonstrates that it is possible to
determine the Balmer decrement to some accuracy from emission line
equivalent widths, the determination of the correction is 
dependent upon the measurement of three
quantities; EW(\ha), EW(\hb), and the flux ratio, 
${\rm F}_{\lambda}(\ha)/{\rm F}_{\lambda}(\hb)$. While the former two will
be observable in strong emission line galaxies at high redshift, the
flux ratio may prove problematic to measure from spectra. However this
ratio is closely tied with the observed optical colours of the
galaxy. In figure \ref{fig:Bcont_colour} we show the distribution of
the continuum fluxes measured at the \ha\ and \hb\ wavelengths
 (${\rm F}_{\lambda}(\ha)/{\rm F}_{\lambda}(\hb)$) against the restframe
 $g-r$ colour as measured from the SDSS fibre spectrum within the
 full SN(\ha,\hb) sample. A linear fit to the correlation in this
 figure returns
\begin{equation}
\log\left({\rm F}_{\lambda}(\ha)/{\rm F}_{\lambda}(\hb)\right)=-0.26+0.39(g-r)_{\rm spec},
\end{equation}
 with a standard deviation of $\sigma \sim 0.015$ dex around this
 relation. We use the rest-frame $g-r$ colour as a proxy for stellar
 continuum, but other colours such as $r-i$ provide similar constraints.
 In the case of low-S/N spectra, the rest-frame colour could be from SED fitting to broad-band magnitudes.

Thus, in summary, the Balmer decrement for a low-resolution,
strong-emission-line spectrum can be measured from the emission line equivalent
widths and colours alone with;
\begin{equation}\label{eqn:EWhahb}
\log(\ha/\hb)=\log\left(\frac{{\rm EW}(\ha)+4.1}{{\rm EW}(\hb)+4.1}\right)+
   \left(-0.26+0.39(g-r)_{\rm rest}\right),
\end{equation}
with a scatter around this of $\sigma\sim0.05$ dex, or $\sim0.3$ mag
in $A_{V}$, assuming a Galactic extinction law \citep[e.g.][]{ODonnell94}.
For weaker emission line galaxies ((i.e.~EW(\ha)$<30$\AA), a lower
offset ($R\sim3.5$) should be used, as
shown in figure \ref{fig:binEWfix}. However, given the larger
uncertainties and greater dispersion seen at lower EW(\ha), a
correction factor of $R\sim4$ can be used for the full sample with a 0.1
dex scatter ($\sim0.7$ mag in $A_{V}$) and an extension to lower
values (i.e~EW Balmer decrement underestimate) due to low EW systems.

One final note on this relation: as seen in figure \ref{fig:EW_HaHb},
there is a strong bias in the sample of Balmer decrement  with other
galaxy properties, as discussed in 
detail in several other SDSS papers \citep[see
e.g.][]{Kauffmann03a,Garn10}. Thus, the relation shown above includes a
combination of both galaxy type as well as variation in extinction.
The only way to remove fully this effect is to match pairs of
galaxies in as many property types excluding extinction,
such as done in \citet{Wild11b}. Unfortunately when applied to the
sample here, it was found that the range in extinction was not large
enough to properly determine the relation. However, even given these
uncertainties, this relation should still hold at several redshifts as high
attenuations are on average
associated with high gas masses, and thus high star formation rates
and similar underlying continua at all redshifts.

\section{Stellar Absorption Effects on the Emission Lines}
\label{sec:offset}

One issue with the previous section is that we assume throughout that
the stellar-continuum subtracted emission line fluxes within the
MPA/JHU database are correct. While the overall fits to the stellar
continuum are impressively good with a median  $\chi^2 = 1.01$ per pixel across the
sample, there are appear to be remaining issues around the Balmer
lines. In the following we concentrate on the SDSS DR7 Balmer emission-line fluxes
corrected for the underlying stellar absorption features from the MPA/JHU
database, and explore their uncertainties using the known intrinsic
values and commonly used attenuation and extinction laws.

\subsection{The issue with \hb}
When considered alone, the ratio of \ha/\hb\ cannot indicate problems with
the measurement of the lines involved unless it is significantly below the
expected value of the unattenuated ratio. This is because the
larger values of the emission line
ratio can be caused by attenuation by intervening dust, with the intrinsic ratio dependent
the emitting gas density and temperature (as discussed
in section \ref{sec:balmer}). However by
examining several of the Balmer lines at once these dependencies can be accounted
for.

\begin{figure}
\includegraphics[width=\hsize]{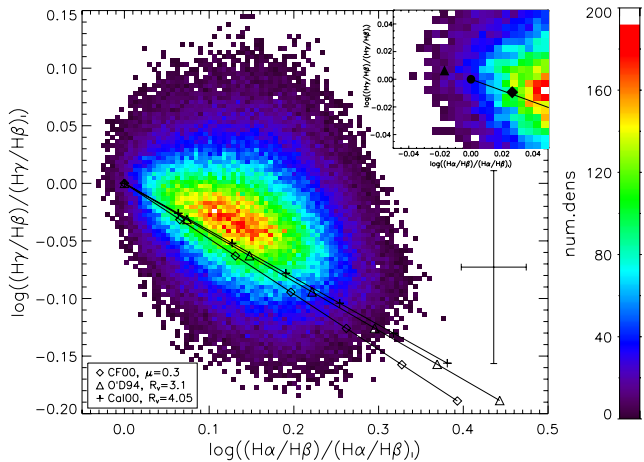}
\caption{The distribution of \hg/\hb\ ratios
versus \ha/\hb\ ratios for the SDSS SN(\ha,\hb,\hg) sample, with the number density of each pixel
indicated by the bar on the right and pixels with less than 10
galaxies not shown. Both axes are normalised to their respective Case
B ratios with $(\ha/\hb)_{\rm I}=2.86$ and $(\hg/\hb)_{\rm I}=0.468$.
The median uncertainty of the sample is indicated by the error bars on
the right. Overplotted are the variation expected from three different
attenuation laws; \citet[][CF00, diamonds]{Charlot00},
\citet[][O'D94, triangles]{ODonnell94}, and\citet[][Cal00, plus
signs]{Calzetti00} (see text). Each symbol represents a step of 0.5 in
$A_{V}$, up to $A_{V}=3$. {\it Inset}: A zoomed in version of the figure, showing
 the position of the zero point when the intrinsic ratios ($(\ha/\hb)_{\rm
    I}$ and $(\hg/\hb)_{\rm I}$) are taken to be at T=5,000K (diamond:
  3.04, 0.458), at 10,000K (circle: 2.86, 0.468), and at 20,000K
  (triangle: 2.75, 0.475). The solid line shows the variation of the
  ratios with the \citet{Calzetti00} attenuation law.}\label{fig:HaHbHgHb}
\end{figure}

Figure \ref{fig:HaHbHgHb} shows the  variation of the \hg/\hb\ ratio
against the \ha/\hb\ ratio for the SN(\ha,\hb,\hg) SDSS sample. Both
ratios have been normalized to their Case B, T=10,000K,
$n_{e}=100$\pccm\ intrinsic ratios ($(\ha/\hb)_{\rm I}=2.86$,
$(\hg/\hb)_{\rm I}=0.468$). Immediately obvious in this figure is the
offset of the sample from the
zero point, indicating most SDSS galaxies undergo some attenuation
(as seen in the previous plots), with the correlation
between the two ratios as expected from the
reddening laws applied to the intrinsic ratio. 

The three different lines overplotted show the effect of three
different attenuation laws commonly assumed in the analysis of
galaxies. For all three lines, the symbols indicates steps of 0.5 in
$A_{V}$, up to $A_{V} = 3$.

The \citet{ODonnell94} law (O'D94) is an updated version of the
\citet{Cardelli89} fit to the average extinction law in the Galaxy,
thus least representative of the integrated emission from the SDSS galaxies, which will suffer
attenuation due to the mixture of emitting sources and absorbing
medium. However, as discussed in \citet{Kennicutt09} and can be seen in
figure \ref{fig:HaHbHgHb}, the use of a foreground dust screen with
galactic extinction is indistinguishable from the other laws,
especially given the uncertainty within the SDSS sample.
We assume a total to selective $V$-band extinction of
$R_{V}=3.1$, the average value in our galaxy. 
 
The \citet{Calzetti00} attenuation law (Cal00) was obtained from the
continuum and Balmer decrement of local actively star-forming galaxies, thus
matching the high EW(\ha) galaxies in the sample. Note that as only
ratios are analysed here, the difference between the colour excess
($E(B-V)$) of the stellar continuum and nebular lines noted by
Calzetti et al.\ is effectively scaled out. The
$R_{V}$ used here is 4.05, as given by \citet{Calzetti00} from
the comparison of the observed infrared flux to that predicted from
the obscuration of the optical-ultraviolet light.

The \citet{Charlot00} attenuation law (CF00) is a more simple, empirical law put
forward to allow for the different colour excesses and attenuations
observed by \citet{Calzetti00} between the nebular emission lines and
stellar continuum. It breaks the attenuation into two components; the
`diffuse ISM' component that describes the effective obscuration
 of all stars in a galaxy by the diffuse dust, and the `birth
cloud' component that describes the additional extinction suffered by the \hii\
regions from which the nebular emission lines arise, giving
\begin{equation}
\frac{A_{\lambda}}{A_{V}}=\mu(\lambda/\lambda_{V})^{-0.7}
 +(1-\mu)(\lambda/\lambda_{V})^{-1.3},
\end{equation}
where $\lambda_{V}=5500$\AA.
The exponent of $-0.7$ for the diffuse ISM was empirically derived by
\citet{Charlot00} with a comparison of nearby galaxies, while the
$-1.3$ exponent for the birth clouds was chosen to match the
extinction within our own Galaxy. The parameter $\mu$ indicates the
fraction of the attenuation suffered by the nebular lines by each
component. We assume $\mu=0.3$ here, as used by \citet{Wild07} and
\citet{Wild11a} in their analyses of SDSS galaxies.

There are three things of note to take from figure \ref{fig:HaHbHgHb}:
One, given enough precision in the data, the SDSS galaxies should be able
to distinguish between these three different attenuation laws, and
provide an answer on which law is best (or least bad) to apply to an
ensemble of galaxies. Such work has been done before for comparing galactic
extinction laws using planetary nebulae \citep{Phillips07}. Note that the different $R_{V}$ between
the Cal00 and O'D94 laws is what causes the
difference in expected \ha/\hb\ for the same $A_{V}$, while the CF00
and Cal00 have similar $A_{V}$ as determined from \ha/\hb, but not
from \hg/\hb.
Two, that the scatter of the SDSS galaxies is
large around the three laws, preventing this possibility, though this
scatter is not significant when compared to the median uncertainty as
shown by the error bars. Third, and most importantly, while the slope
of SDSS galaxies matches that given by the attenuation laws, \emph{there is a
systematic offset of the SDSS galaxies when compared to all three
attenuation laws}. This offset is significant, and cannot be explained
by assuming more extreme (and therefore less likely) attenuation laws
or by assuming large values of $R_{V}$, as the zero point itself appears to be
offset.

Neither can this situation be remedied by assuming different values
for the unattenuated Balmer ratios. As discussed in section
\ref{sec:balmer}, the intrinsic Balmer emission-line ratios are
sensitive to the temperature of the ionized gas from which
they arise, and, more weakly, to the density of the gas as well
The inset in figure \ref{fig:HaHbHgHb} shows a
close up of the zero point of figure \ref{fig:HaHbHgHb}, i.e.~galaxies
with little or no attenuation. Over this are
plotted 3 symbols indicating the position where the zero point would
be for 3 different average \hii\ region temperatures; 5,000K, 10,000K
(assumed within the figure),
and 20,000K. All assume Case B ratios, and a typical \hii\ gas density of $n_{\rm
  e}=10^2$\pccm . These 3 temperatures encompass the range of
temperatures expected, with typical solar metallicity \hii\ regions
having T$_{\rm e}\sim$8,000K. What this figure demonstrates is that
the variation in intrinsic Balmer ratios is small relative to both the effects
of attenuation and the observed offset, and that the variation in intrinsic
ratios is in the same sense as that due to the effects of attenuation (as shown by
the \citet{Calzetti00}  law), thus the intrinsic ratio cannot be the cause of the offset.

\begin{figure}
\includegraphics[width=\hsize]{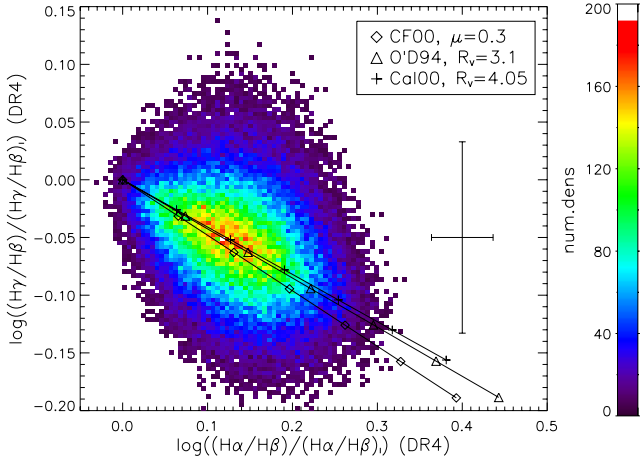}
\caption{The same plot as figure \ref{fig:HaHbHgHb}, except using the SDSS
  DR4 MPA/JHU catalogue. The same SN cuts on \ha, \hb, and \hg\ are
  applied, with $\sim$160,000 galaxies in the sample. }\label{fig:DR4HaHbHgHb}
\end{figure}

A possible cause for this offset can be found when the MPA/JHU DR4
catalogue is examined instead. Using the same SN cuts on \ha, \hb, and
\hg, figure \ref{fig:DR4HaHbHgHb} shows the same plot as figure
\ref{fig:HaHbHgHb} for the DR4 sample. In most ways this figure is
exactly the same (as it should be), except in two respects: the DR4
SN(\ha,\hb,\hg)  sample only has $\sim$160,000 galaxies in the sample
compared to the $\sim$240,000 galaxies in the DR7 sample,
visible in both the number density (colours) and in the scatter,
and the DR4 sample does not have the offset with respect to the attenuation
laws seen in figure \ref{fig:HaHbHgHb}.
%
%
%
%

As mentioned in section \ref{sec:SDSSsample}, the major difference
between the DR4 and DR7 line fluxes from the MPA/JHU catalogues is the
version of the GALAXEV models used for the continuum fits. 
In some respects the difference seen in the figures is surprising, as the median $\chi^2$ of the
fits to the continuum in DR7 is reduced compared to the DR4 fits, from
$\chi^2=1.5$ for DR4 to $\chi^2=1$ for DR7, suggesting a much better
fit to the spectra.

\begin{figure}
\includegraphics[width=\hsize]{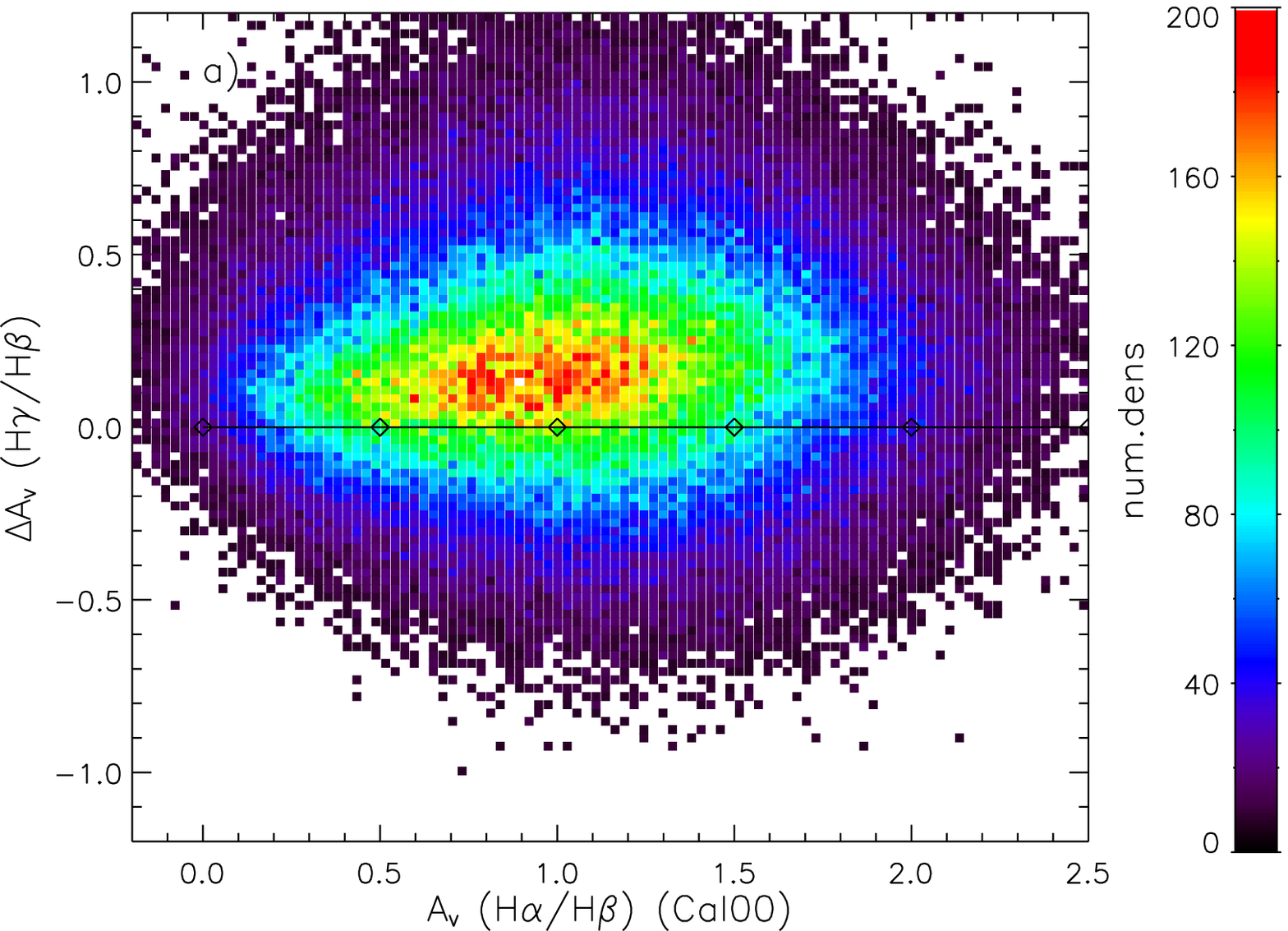}
\includegraphics[width=\hsize]{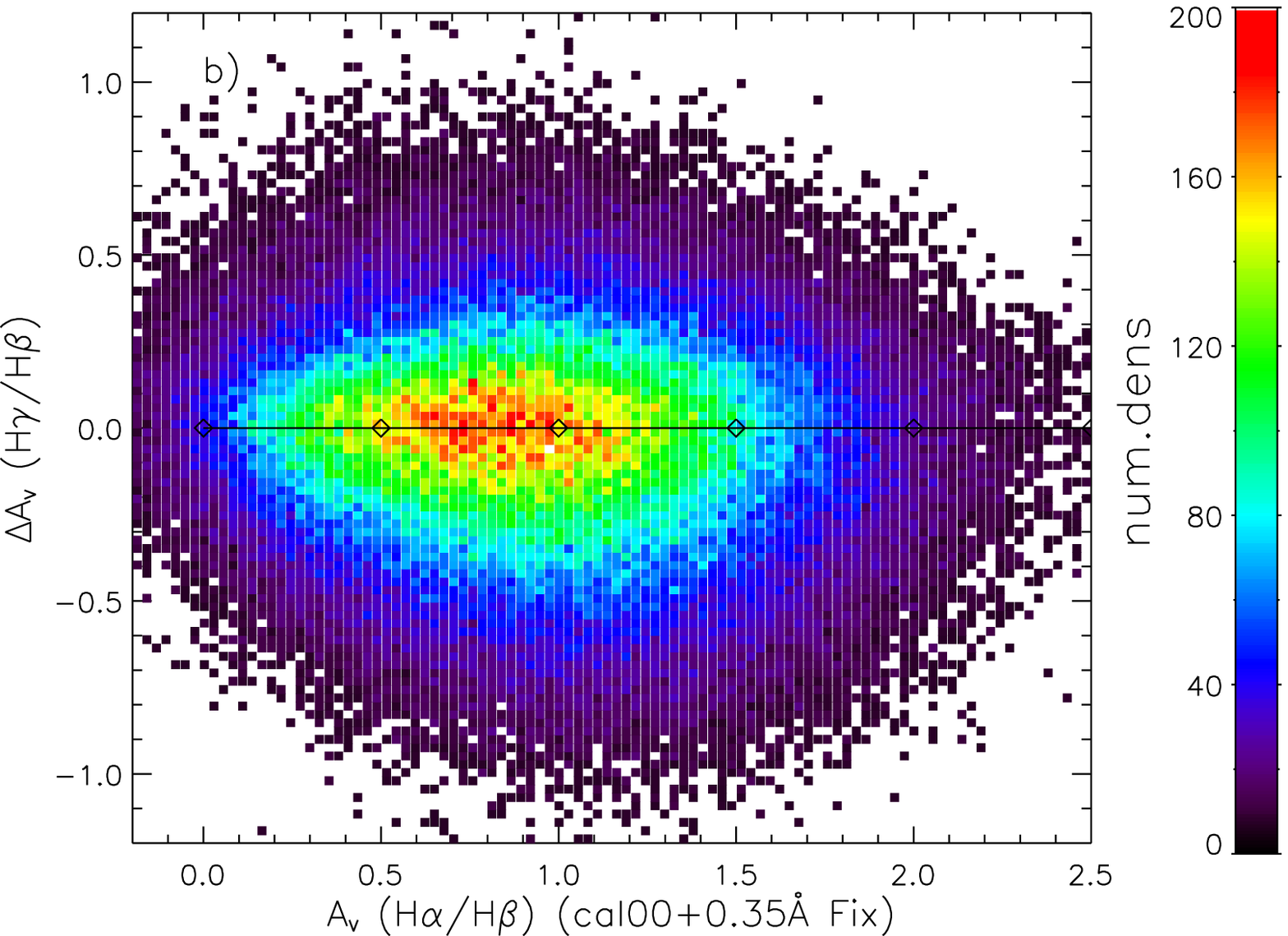}
\caption{
{\bf a)} The same diagram as Figure \ref{fig:HaHbHgHb}, but rotated to $A_{V}$ axes using the
  attenuation law of \citet{Calzetti00}. The $x$-axis is the $A_{V}$
  as measured from the (\ha/\hb) ratio, while the $y$-axis is the
  offset from the expected (\hg/\hb) ratio based on this $A_{V}$. 
{\bf b)}The same as a), but the \hb\ line flux has been
 corrected for under subtraction of the stellar continuum by
  0.35\AA.}\label{fig:DR7_AvHaHb} 
\end{figure}

The issue most likely lies within the \hb\ line region, as an offset in the \hb\ line
flux would also explain the increasing offset at higher (\ha/\hb) between the SDSS
galaxies and the attenuation laws, and this region
has been observed to be mismatched between models and the spectra of
some globular clusters \citep[see e.g.][]{Walcher09,Poole10}. 
This difference in slope is more clearly seen in figure
\ref{fig:DR7_AvHaHb}a where we have rotated and scaled figure \ref{fig:HaHbHgHb}
to the $A_{V}$ plane using the attenuation law of \citet{Calzetti00},
where the offset from the $x$-axis is more clearly seen. Similar
results are seen if another attenuation law is used. In this frame, the $y$-axis is the
offset from the expected (\hg/\hb) ratio based on the $A_{V}$ as
determined by the (\ha/\hb) ratio.

By ``correcting'' for the incorrectly subtracted stellar \hb\ equivalent width we
can fix figure \ref{fig:DR7_AvHaHb}a. This is what we have done in
figure \ref{fig:DR7_AvHaHb}b, where we have added 0.35\AA\ to the
emission line equivalent width (i.e.~``True''
\hb=\hb+0.35F$_{\lambda}$(\hb)). The correction of 0.35\AA\ was
determined by minimizing the offset from the Calzetti law (this value depends
only weakly on the choice of attenuation law). This means that the stellar
absorption equivalent width of \hb\ is systematically underestimated
by 0.35\AA\ in the CB08 continuum fits to DR7 SDSS spectra.

The fact that this 0.35\AA\ underestimation is systematic is
interesting, as we would expect that any error would depend on the
strength of the Balmer features, either measured through the \hd$_{\rm
  Abs}$ or the emission line equivalent widths, 
both of which do correlate with \ha/\hb\ as seen in figure
\ref{fig:EW_HaHb} , yet no correlation is observed. 
What the underlying cause of this systematic underestimation is
unknown, yet it must be taken into consideration when determining the
$A_{V}$ from the Balmer decrement in the SDSS DR7 MPA/JHU
catalogue. It is possible that this issue arises due to the misclassification of
the spectral resolution of the MILES library \citep[as discussed
in][]{FalconBarroso11} in the implementation in the CB08 code, 
but this issue is now known and currently under investigation. This
investigation goes beyond the
scope of the work presented here but we note
that the models used here are early versions of the CB08 library
and these issues are expected to be solved within the to-be-published models. 
When the new $A_{V}$ is calculated from the \ha/\hb\ ratio including
the systematic 0.35\AA\ offset \citep[using e.g.~][law]{Calzetti00}, a
mean difference of $-$0.07 magnitudes is found with the uncorrected
$A_{V}$ estimates, increasing slightly at higher $A_{V}$. This suggests that
previous DR7 $A_{V}$ estimates, such as in \citet{Garn10},  are
overestimated by this value.  

Similarly this 0.35\AA\ correction must be included in our EW Balmer
decrement (equation \ref{eqn:EWhahb}) leading to a new equation,
\begin{equation}\label{eqn:EWhahbfix}
\log(\ha/\hb)=\log\left(\frac{{\rm EW}(\ha)+4.1}{{\rm EW}(\hb)+4.4}\right)+
   \left(-0.26+0.39(g-r) _{\rm rest}\right),
\end{equation}
which more closely matches our expectations of different stellar
absorption equivalent widths between \ha\ and \hb. For weaker emission
line galaxies, the offset would be smaller than 4.1, as shown in
figure \ref{fig:binEWfix} and discussed at the end of section
\ref{sec:EWBalmer}. 

\subsection{The issue with \hd}

\begin{figure}
\includegraphics[width=\hsize]{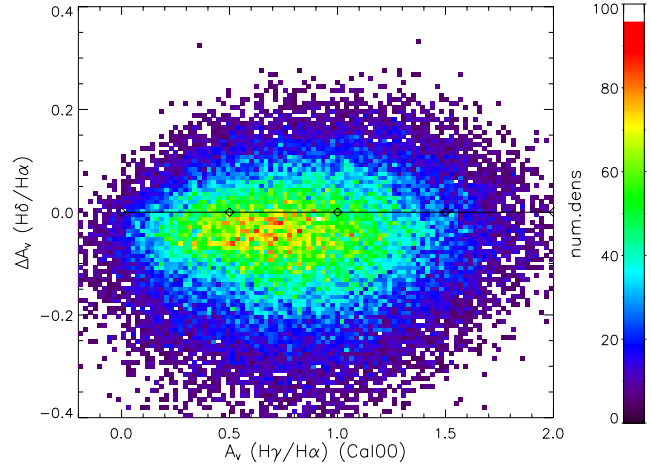}
\caption{The distribution of \hd/\ha\ versus \hg/\ha\ for the
  SN(\ha,\hb,\hg,\hd) sample of SDSS DR7 galaxies ($\sim$121,00
  objects). As in figure \ref{fig:DR7_AvHaHb}, the figure is rotated
  and scaled to the $A_{V}$ plane as 
  determined from the \citet[][]{Calzetti00} law, indicated by the
  straight line. The number of 
  galaxies in each pixel is indicated by colours (as labelled), with
  pixels with less than 5 galaxies not shown.}\label{fig:AvHgHa}
\end{figure}

While examining the issue in \hb, a similar issue was found for \hd\
that we present here as a curiosity. When \hd/\ha\ versus \hg/\ha is
plotted, using the SN(\ha,\hb,\hg,\hd) sample of SDSS galaxies and
avoiding the problematic \hb\ line, the tight 
correlation between these two ratios, matching closely that expected
from the attenuation by dust. However, upon closer examination, a systematic offset is
observed between the galaxies and attenuation laws. 
As in the previous
subsection on \hb, the offset is clearer when rotated to the
$A_{V}$ plane, which is what we show in figure \ref{fig:AvHgHa}, where the $x$-axis
is the $A_{V}$ determined from \hg/\ha\ using the \citet{Calzetti00}
law, with the $y$-axis the offset from this $A_{V}$ when determined
from the \hd/\ha\ ratio. The 
median offset is $\sim-0.05$ for most of the $A_{V}$ range shown here,
meaning that the  $A_{V}$ determined from \hd/\ha\ is lower than that
determined from \hg/\ha. The offset is seen for all attenuation laws
considered here. 

More importantly, the offset is also observed in the DR4
sample, though with greater uncertainty due to low number statistics. 
 As with the \hb\ line there
appears to be no correlation of the offset with emission line
equivalent widths, or stellar age determinants like H$\delta_{\rm
  Abs}$ or the D$_{\rm n}4000$ index \citep[see][ for
definitions of these indices]{Kauffmann03a}. Neither does it appear to
be correlated with \ha/\hb\ or other emission line or attenuation tracers.
Thus, due to the lack of difference between DR4 and DR7, the strong EW bias
of the sample (as shown in figure \ref{fig:EWHa_wcuts}), and the fact
that the uncertainty is dominated by the \hd\ line, it is still not
known what exactly causes this offset. It is most likely an issue due
to the underlying continuum, but an investigation into the stellar
models goes beyond the scope of this work. Thus we present this issue for now as a
curiosity and a cautionary note of the level of systematic
uncertainties in determining weak line fluxes from the SDSS sample. 

\section{Conclusion}

We have examined the possibility of using equivalent widths of the
Balmer emission lines to determine the Balmer decrement, and hence
attenuation, of a galaxy. Using the Sloan Digital Sky Survey we were
able to determine a statistically representative relation between the
continuum-subtracted Balmer emission line flux ratio and the
equivalent
widths (EW) of the Balmer emission-lines combined with a rest-frame
 colour, correcting for the effects of the stellar
absorption features:
\begin{equation}
\log(\ha/\hb)=\log\left(\frac{{\rm EW}(\ha)+4.1}{{\rm EW}(\hb)+4.4}\right)+
   \left(-0.26+0.39(g-r)\right),
\end{equation}
for galaxies with EW(\ha)$\gapprox 30$\AA, with a scatter of
$1\sigma\sim0.06$ dex, or 0.4 mag in $A_V$, indicating the possible
variation for individual objects. For galaxies with
EW(\ha)$<30$\AA\ smaller correction factors (3.5 for EW(\ha), 3.8 for
EW(\hb)) should be used. However, given the scatter at low EW values, the equation
above can be used above for all galaxies allowing for a much greater
uncertainty in the final Balmer ratio or $A_V$ determined.

In addition, by comparing the Balmer decrement (\ha/\hb\ versus
\hg/\hb) we discovered that the \hb\ emission line equivalent width
(and hence flux) is underestimated by $~0.35$\AA\ in the
JHU/MPA DR7 SDSS database, due to an issue in the \hb\ region of the
2008 version of the Charlot \& Bruzual stellar population synthesis
code GALEXEV. This leads to an overestimation of the attenuation of
the SDSS galaxies of 0.07 magnitudes in $A_V$ assuming a
\citet{Calzetti00} attenuation law.

Finally, we also discovered a strange offset in the \hd\ emission line
fluxes observable in both the DR4 and DR7 releases of the MPA/JHU
database which we present both as a curiosity and as a warning on the
underlying issues in interpreting weak-line emission lines in
a statistical sample

\section*{Acknowledgements}
BG would like to thank V.~Wild for providing her PCA principal component
values available and very helpful discussions, and the authors would
like to thank the referee for helpful comments.

Funding for the SDSS and SDSS-II has been provided by the Alfred
P. Sloan Foundation, the Participating Institutions, the National
Science Foundation, the U.S. Department of Energy, the National
Aeronautics and Space Administration, the Japanese Monbukagakusho, the
Max Planck Society, and the Higher Education Funding Council for
England. The SDSS Web Site is \url{http://www.sdss.org/}. 

The SDSS is managed by the Astrophysical Research Consortium for the
Participating Institutions. The Participating Institutions are the
American Museum of Natural History, Astrophysical Institute Potsdam,
University of Basel, University of Cambridge, Case Western Reserve
University, University of Chicago, Drexel University, Fermilab, the
Institute for Advanced Study, the Japan Participation Group, Johns
Hopkins University, the Joint Institute for Nuclear Astrophysics, the
Kavli Institute for Particle Astrophysics and Cosmology, the Korean
Scientist Group, the Chinese Academy of Sciences (LAMOST), Los Alamos
National Laboratory, the Max-Planck-Institute for Astronomy (MPIA),
the Max-Planck-Institute for Astrophysics (MPA), New Mexico State
University, Ohio State University, University of Pittsburgh,
University of Portsmouth, Princeton University, the United States
Naval Observatory, and the University of Washington.  

\bibliographystyle{mn2e}

\end{document}